\documentclass{llncs}
\usepackage{amsmath,amssymb,graphicx,epsfig}
\usepackage{algorithm}
\usepackage{algorithmic}
\usepackage{comments}
\usepackage{subfigure}
\usepackage{luca}
\usepackage{macros} 

\author{Pritam Roy \\ pritam@ee.ucla.edu}

\institute{Computer Science and Electrical Engineering Department \\ 
University of California,Los Angeles, USA
}

\title{Interface Building for Software by Modular Three-Valued Abstraction Refinement}

\begin{document}
\maketitle
\begin{abstract}
Verification of software systems is a very hard problem due to the large size of program state-space.
The traditional techniques (like model checking) do not scale; since they include the whole state-space by inlining the library function codes.
Current research avoids these problem by creating a lightweight representation of the library in form of 
an {\em interface graph}  (call sequence graph).
In this paper we introduce a new algorithm  to compute a safe, 
permissive interface graph for C-type functions.
In this modular analysis, each function transition is summarized following three-valued abstraction semantics.
There are two kinds of abstraction used here.
The global abstraction contains predicates over global variables only; however the local abstraction 
inside each function may also contain the local variables.
The abstract summary needs refinement to guarantee safety and permissiveness.
We  have implemented the algorithms in TICC tool and compared this algorithm with some 
related interface generation algorithms. 
We also discuss the application of interface as an offline test-suite. 
We create an interface from the model program (specification)  and 
the interface will act as a test-suite for the new implementation-under-test (IUT).
\end{abstract}
\section{Introduction}

Verification of software systems is a very hard problem due to the large size of program state-space. 
Most software programs contain library functions and these kind of functions are examples of
\emph{open systems}.
The verification of such open systems becomes infeasible due to two main problems. 
Firstly, in order to verify a given program one needs to \emph{inline} the library function code and it increases the space complexity of the verification algorithms.
Current formal techniques like model-checking can not handle the large state-space generated 
from the  program variables.
The second option is to verify the library functions a priori so that there is no need to inline them.
For this purpose, most of the time a small code containing a sequence of library  functions calls(called \emph{client}) is written.
The client code invokes the library functions to close the open system. 
The library functions are impossible to verify in the absence of exhaustive client program. 
Hence most of the verification approaches plug-in a client code to close the open-system.

\subsection{Interface and Properties}

 The current research \cite{FSE05,ACMN05,CAV07b} avoids these two problems by applying  \emph{modular verification} techniques which builds a small call sequence graph, called \emph{interface} representing union of all client programs. 
%
The  interface contains all possible call sequences which leads the library to error or illegal states. 
Similarly, the interface should contain all possible call sequences which avoids the error states. 
Henceforth constrains on the use of the library function calls from outside and the user can distinguish
the legal call sequences from the illegal ones by simply looking at the interface. 
 There are two immediate benefits of using the interfaces. 
Firstly, these interfaces are light-weight representation of the libraries and the implementation of the library functions can be replaced by the interface.
Secondly, the interfaces can be constructed without the help of any client program.
The interface should be \emph{safe} i.e. all illegal call sequences (which leads the library to the error states) will be present in the interface.
 The interface graph should be \emph{permissive}  i.e.  all legal sequences will be present in
 the interface.


\subsection{Related Work}

However, there are some challenges in building succinct interfaces.
The interface size can become exponential in terms of number of 
variables.
A symbolic representation and abstraction techniques partition the state-space into a small number of regions where every region represents one node of the interface graph. 
Some researches  apply these abstraction and symbolic techniques to obtain a small but safe and permissive interface.
 
 The work by Alur et. al. (~\cite{ACMN05}) uses {\em Angluin's learning algorithm} L* to create 
 an interface.
The algorithm learns the interface language by asking membership and equivalence queries to teacher
(here program).
The generated interface is  safe and minimal; but not permissive. 
To handle big case studies predicate abstraction has been used, however the user need to 
provide the predicates.
There is no automatic abstraction refinement.
The algorithm returns minimal size interface if the algorithm is not hit by timeout.
Experimental results show that  even in small examples timeout occurs.
The CEGAR approach by Henzinger et. al. (~\cite{FSE05}) creates a safe and permissive interface.
The size of the interface can be big enough depending on the chosen counter-example. 
The direct approach by Beyer et. al. (~\cite{CAV07b}) creates an interface which is safe and permissive.
This approach does not use abstraction and hence the interface can become very large.



\subsection{Contribution} 

Unlike the related work, our work can also be used in unstructured or non-object oriented 
(C style) functions.
In an object-oriented framework every class variable is accessible to every class method
and can be a global variable to the class method.
Instead we assume that each function may contain several local variables in addition to those
global variables.
Hence, we have {\em more general platform} to compute interface. 
Each of these functions  can also have several sequential updates of variables, call to other functions
even recursive calls to themselves.
However, we compute the interface including only functions accessible to the user level. 

In the first stage of three stage algorithm, every C library function is parsed by CIL (C Intermediate Language)\cite{CIL} and converted into TICC \cite{TICC-tool} input language.
This language syntax is similar to the guarded-update language.
We have implemented the next two stages in this Multi-valued Decision Diagram \cite{MDD90}-based symbolic tool TICC.  
The second stage computes the transition summary of each function. 
This modular algorithm handles each function separately including local variables within the scope.
However,  the space complexity of function summary becomes a bottleneck in order to compute big functions which may contain large number of guarded-updates. 
Hence, we employ  three valued abstraction refinement schemes in addition to symbolic techniques.
The abstraction in summarization ensures small size; whereas successive refinement of the 
abstract states fine tune the abstraction to obtain the safety and permissiveness.
In the last stage, an interface graph is built from the abstract set of states. We show different
stages of building a symbolic safe and permissive interface in the following example. 
\begin{example}[Motivating Example]
Figure~\ref{intstackcode} defines a stack data-type $stackT$ and two functions $push$ and $pop$.
The data type $stackT$ has an array of integers $el$ of size $MAX$ and an integer showing the $top$
of the stack. 
The function $pop$ returns error when the stack is empty i.e. top is zero. The function $push$
returns error if the top is equal to $MAX$. 
Otherwise copies the input value $sd$ into the $el$ array at address $top$. 
The $top$ is incremented later.
Figure~\ref{intstackrule} shows how the C code is converted into guarded-update rule in the next stage.
The global variable {\em err} denotes the error in the library and the library goes to error state
when {\em err} is set to 1.
Figure~\ref{intstackinterface} shows the interface graph from the set of rules.
The initial state of the interface graph is {\em state 1} where the stack is empty.
A call to pop function from the initial state will move the library into an {\em ERROR} state.
Similarly calling push form state $3$ will be an error due to full stack. 
We can note that the interface can create many legal as well as illegal sequences of stack functions. 
To check each of them we otherwise need a set of client programs.
\begin{figure}[htb]
\centering
\subfigure[Code]{
\includegraphics[scale=0.3]{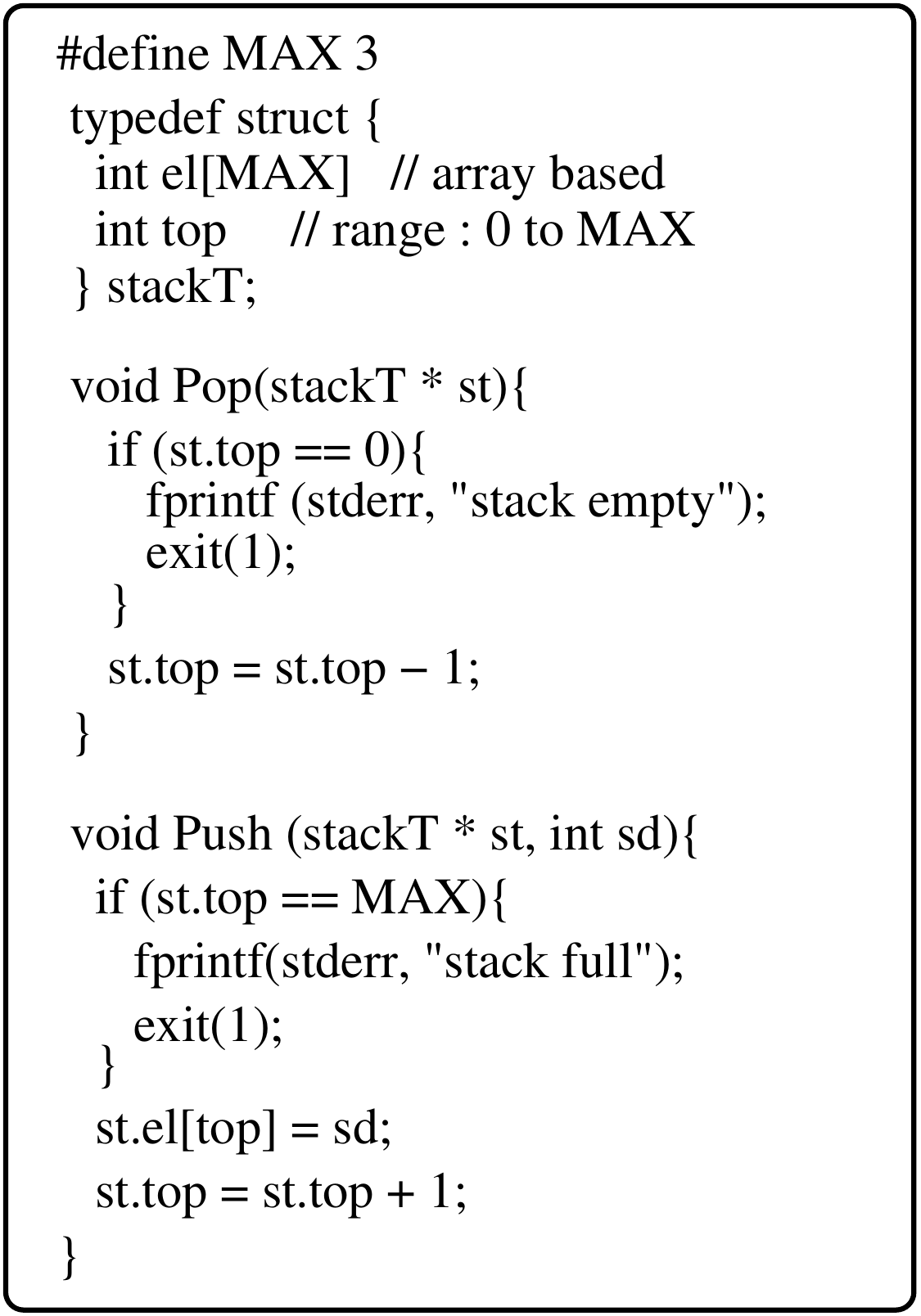}\label{intstackcode}
}
\subfigure[Rules]{
\includegraphics[scale=0.31]{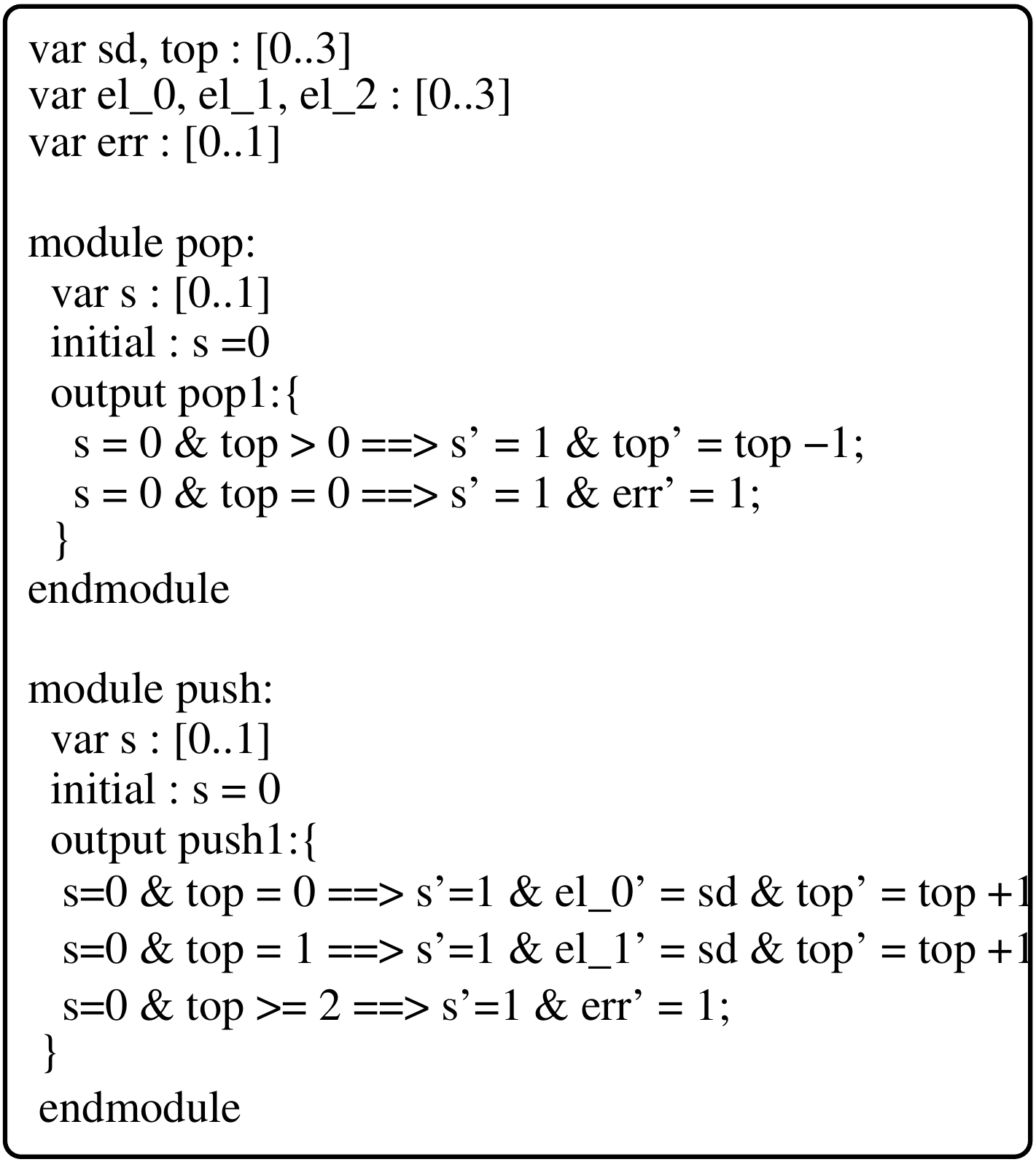}\label{intstackrule}
}
\subfigure[Rules]{
\includegraphics[scale=0.31]{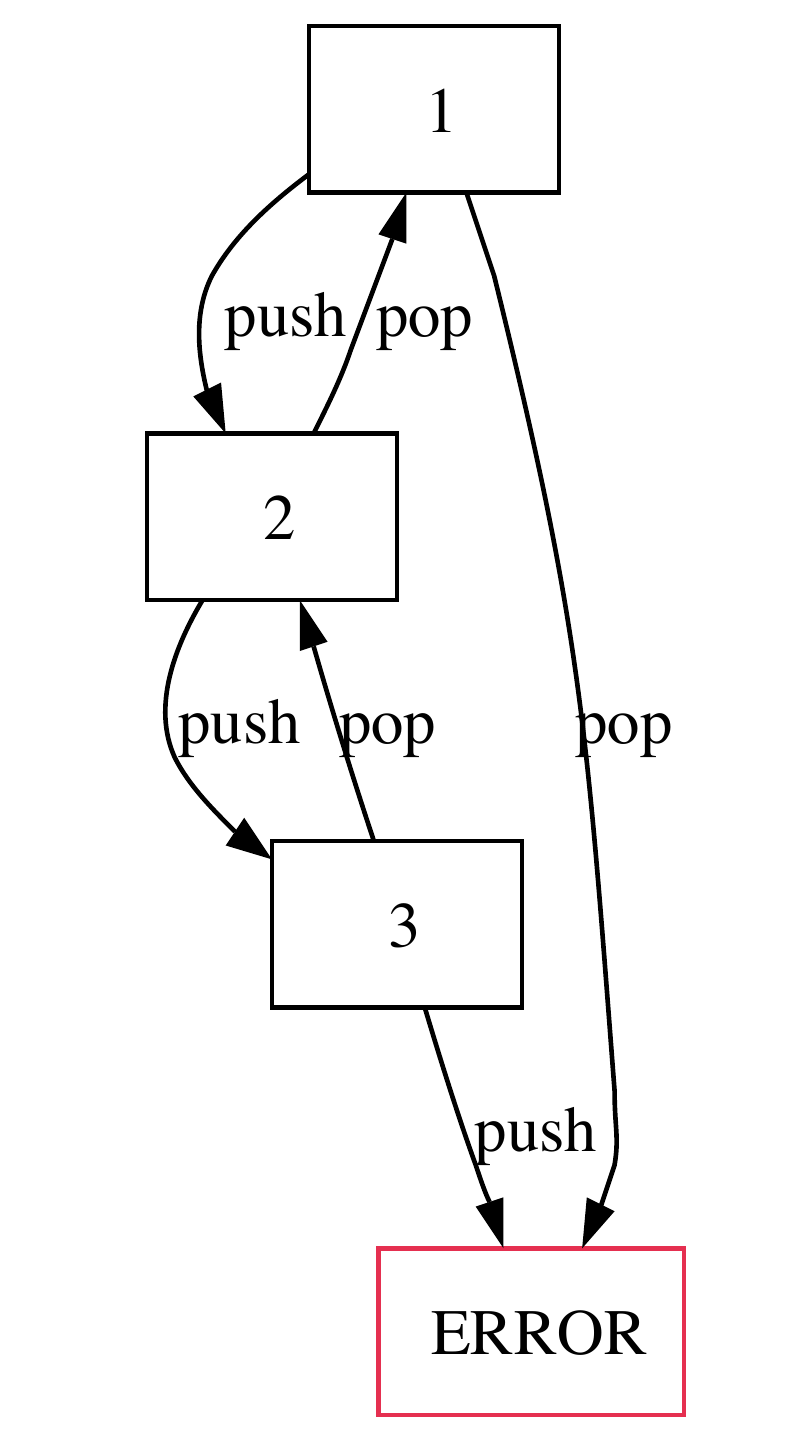}\label{intstackinterface}
}
\caption{Stack Example}\label{intstack}
\end{figure}
\end{example}

 Finally we discuss the applications of the safe and permissive interface graph.
 Firstly, any given client program can immediately verify with the help of the interface graph whether the function call sequence in the client leads the library to some error states.
Secondly, the interface can actually provide an offline test-suite for a set of functions.
Often the source of the library is unknown; however one can create a model program from the
available documentation of the functions.
The interface graph obtained from the model program can be used to test the
implementation-under-test (IUT).

\section{Preliminary Definitions}\label{def}
In this section we provide preliminary definitions and the background work.
\subsection{A Transition System Model for Libraries}
A software library module $Lib = (F_G,V_G, E, I)$ contains a set of functions $F_G$ and a set of 
global variables $V_G$. 
The global variables $V_G$ constitute variables declared outside any of the functions in $F_G$.
The {\em global state space} $S_G$ can be defined with respect to different valuations of global variables $V_G$. 
The variable $err \in V_G$ is a special global variable in $Lib$ which can take two values $0$
and $1$.
The library reaches an error set  $E \subseteq S_G$ when the global variable $err$ is set to $1$. 
Moreover, the error set is a sink set of the library.
The initial configuration of the library is given by set $I \subseteq S_G$.

Each function $f \in F_G$ also contains a set of local variables $V^f_L$.
The {\em scope} of any local variable $v \in V^f_L$ is function $f$.
There is a special local variable, called {\em s}, in $V^f_L$ which corresponds to the relative location
in the function with respect to the first location. 
For a function $f$, all variables $V^f$ can be given as $V^f_L \cup V_G$ and 
 {\em function state-space} $S_f$ can be defined with respect to different valuations $V^f$. 
 We note that each global set $s_G \in S_G$ is a non-empty subset of $s_G \subseteq S_f$
 function state-space. 
 The {\em initial local state set} $I^f_L \subseteq S^f$ denotes the entry point to the function $f$. 
All variables of the library $Lib$ is denoted by $V$ and is given by $V {:=} V_G \cup \cup_{f \in F_G} V^f_L$.
The {\em total state-space} $S$ can be defined with respect to different valuations of all variables $V$.   

Each function $f \in F$ contains some number (say $k$) of guarded-update rules. 
For  $i$-th such rule, its condition part $i.guard \subseteq S_f$ can be given as a set of function states, and the assignment part $i.update \subseteq S_f \times S_f$ can be given as the set of transitions.
For a set $X \subseteq S_f$, $i.update (X) : S_f$ denotes the next state of X in the $i-th$ update rule.
The conditional transition of rule $i$ given as 
\[
i.trans { := }  \set{(s_1, s_2) \in S_f \times S_f \mid s_1 \in i.guard, s_2 \in i.update(i.guard)}.
\]
The transition relation $Trans^f \subseteq S_f \times S_f$ can be given as the union of rules corresponding to the function $f$ i.e. $Trans^f {:=} \cup_{i=1 \ldots k}  i.trans$.
We will use $Trans^f (t) \subseteq S_f$ to denote the successor set of state $t \in S_f$.  

For a binary relation $\Join \in \set{=,\leq,\geq}$ and a state-space $S$, the set 
$S\mid_{v \Join a}$ denotes the set where the value of a variable $v$ related to value $a$ with relation $\Join$. 
For a set $X \subseteq S_f$, we define $support(X) \subseteq V_f$  as the set of variables whose
value change result in a value change of $X$.
Formally we can write, 
\[
support(X) := V^f \setm \set{v \in V^f \mid \forall s,s' \in S_f. s =_{v} s'  \im s\in X \iff s' \in X}
\]
where $s\ =_{v}\ s'$ implies that $s=s'$ except for a variable $v \in V^f$.
 Interface graph is an input-enabled interface automata.
 Given a Library $Lib = (F_G,V_G, E, I)$ and global state-space $S_G$,
 we can define  {\em interface-graph} or call sequence graph as  $IG = \langle N, T, T_{e},In, Er \rangle$ where, 
 \begin{itemize}
 \item the nodes $N \subseteq 2^{2^{S_G}}$ correspond to the set of states,
 \item the set $In \subseteq N$ denotes the initial nodes corresponding to $I$, 
 \item the set $Er \subseteq N$ denotes the error nodes corresponding to $E$,
 \item the set $T \subseteq N \times F_G \times (N\setm Er)$ denotes  good transitions.
  \item the set $T_{e} \subseteq N \times F_G \times Er$ denotes erroneous transitions.
 \end{itemize}

\subsection{Three Valued Abstraction} 
For a library $L=(F_G,V_G)$, a function $f \in F_G$ and a function state-space $S_f$, 
an {\em abstraction\/} $R  \subs 2^{2^{S_f} \setm \emptyset}$ is defined such that
each abstract state (or region) $r \in R$ is a non-empty subset $r \subs S_f$ of
concrete states.   
We require $\bigcup R = S_f$.  
For subsets $T \subs S_f$ and $U \subs R$, we write: 
\[
  \concr{U} = \textstyle \bigcup_{u \in U} u 
  \qquad
  \absm{T}  = \set{r \in R \mid r \inters T \neq \emptyset} 
  \qquad
  \absM{T}  = \set{r \in R \mid r \subs T}
\]
Thus, for a set $U \subs R$ of abstract states, $\concr{U}$
is the corresponding set of concrete states. 
For a set $T \subs R$ of concrete states, $\absm{T}$ and 
$\absM{T}$ are the set of abstract states that constitute over and
under-approximations of the concrete set $T$. 
We say that the abstraction $R$ of a state-space $S_f$ is 
{\em precise\/} for a set $T \subs S_f$ of states if 
$\absm{T} = \absM{T}$. 

\subsection{$\mu$-Calculus}
We will express our algorithms for solving reachability on the function state
space in $\mu$-calculus notation \cite{EmersonJutla91}. 
Consider a procedure $\gamma: 2^{V^f} \mapsto 2^{V^f}$, monotone when $2^{V^f}$ is
considered as a lattice with the usual subset ordering. 
We denote by $\mu Z. \gamma (Z)$ (resp.\ $\nu Z. \gamma (Z)$) 
the {\em least\/} (resp.\ {\em greatest\/}) {\em fix-point\/} of
$\gamma$, that is, the least (resp.\ greatest) set $Z \subs V$ such 
that $Z = \gamma (Z)$. 
As is well known, since $V$ is finite, these fix-points can be computed
via Picard iteration: 
$\mu Z . \gamma (Z) = \lim_{n \go \infty} \gamma^n (\emptyset)$ and 
$\nu Z . \gamma (Z) = \lim_{n \go \infty} \gamma^n (V)$. 

\subsection{Predecessor Operators}
For a library function $f$ and a function state-space $S_f$,  
we define the  {\em one-step predecessor operator\/}
$Pre^{f,1} : 2^{S_f} \mapsto 2^{S_f}$ as follows, for all $Y \subs S_f$: 
\begin{equation}\label{pref}
  Pre^{f,1} (Y) = \{ x \in S_f \mid Trans^f (x) \cap Y \neq \emptyset\}
\end{equation}
We define the  {\em multi-step predecessor operator\/}
$Pre^{f,*} : 2^{S_f} \mapsto 2^{S_f}$ as follows, for all $Y \subs S_f$: 
\begin{equation}\label{pref-global}
  Pre^{f, *} (Y)  = \{ s \in S_f \mid  s \cap (\mu X. (Y \cup Pre^{f,1}(X))) \neq \emptyset\}
\end{equation}

Intuitively, the set $Pre^{f,*}(X)$ consists a subset of $S_f$ from which one can reach to $X$ by applying zero or more transitions within the function $f$ by applying rules one after another.

For the abstract state space $R$, we introduce abstract versions of $Pre^{f,R}_\cdot$. 
As multiple concrete states may correspond to the same
abstract state, we cannot compute, on the abstract state space, 
a precise analogous of $Pre^{f,R}_\cdot$. 
We define two abstract operators: the {\em may\/} operator $Pre^{f,R}_m: 2^R \mapsto 2^R$, 
which constitutes an over-approximation of $Pre^f$, 
and the {\em must\/} operator $Pre^{f,R}_M: 2^R \mapsto 2^R$, 
which constitutes an under-approximation of $Pre^f$ \cite{dAGJ-lics04}. 
We let, for $U \subs R$: 
\begin{align}
  \label{eq-abstract-cpre}
  Pre^{f,R}_m(U)  & = \absm{Pre^{f,*} (\concr{U})} &
  Pre^{f,R}_M(U) & = \absM{Pre^{f,*} (\concr{U})} . 
\end{align}
%
The fact that $Pre^{f,R}_m$ and $Pre^{f,R}_M$ are over and
under-approximations of the predecessor operator is made
precise by the following observation: for all $U \subs R$ we have
\begin{align}
  \label{eq-comp-pre}
  \concr{Pre^{f,R}_M (U)} \subs Pre^{f,*} (\concr{U}) \subs \concr{Pre^{f,R}_m(U)}
\end{align}.
For an integer $k \geq 1$ and function state-space $S_f$, 
we recursively define the {\em k-step  post operator\/}
$Post^{f,k} : 2^{S_f} \mapsto 2^{S_f}$ as follows, for all $X \subseteq S_f$: 
\begin{align}\label{post-onestep}
  Post^{f,1} (X) & =  \cup_{x \in X}\ Trans^f (x)\\
  Post^{f,k} (X)  & =  Trans^f (Post^{f-1,k}(X))
\end{align}

For an abstract state space $R \subseteq 2^{2^{S_f}}$, we define the {\em abstract post operator\/}
$Post^{f,R}_m : 2^R \mapsto 2^R$ as follows, for all $X \subseteq R$: 
\begin{equation}\label{post-global}
  Post^{f,R}_m (X) =  \{ r \in R \mid r \cap\ Post^{f,k} (I^f_L \cap (\concr{X})) \neq \emptyset\}
\end{equation}
where $k$ is the smallest integer to satisfy $Post^{f,k+1}(I^f_L \cap (\concr{X})) = \emptyset$.
Intuitively, the condition implies that no new states are added in the $k+1$-th iteration, hence 
the last updated value when $f$ returns can be obtained by applying $Post^{f,k}$ to 
a subset of $\concr{X}$ corresponding to the function's initial state set $I^f_L$.

\section{Translation from C to Guard-Update Rules}
In this section we discuss our procedure to convert C functions into the "sociable interface automata"
\cite{frocos05} format.
This format is contains several guarded-update rules and is the input format of our symbolic tool TICC.
In our work the front-end and back-end are separate.
Hence one only need a different front-end to parse functions from any other language (like Java/C++)
to generate the TICC input format models.
The next stages of the algorithm can reuse the out tool TICC to build interface graphs.

The C functions are fed into CIL\cite{CIL} tool which parses C source code and 
returns the control flow graph.
The control flow graph contains block structure as nodes and the conditions as the transitions.
We have modified the control flow graph for each function into set of guarded-update rules.
The conditions are represented as guards and the assignments are represented as updates.
The special local variable $s$ defines the location of current block.
For a variable $v$, the primed variable $v'$ denotes the $v$ in the next sequential step.  
When the translator encounters a critical error condition  (e.g. call to $exit(1)$) in the
control flow graph; the global variable $err$ is set to 1 in the translated library.  

\begin{itemize}
\item Control Flow Structures: The C source like "if (a =0) \{b=0;\} else \{b=1;\}" is converted into the following rules:
\begin{align*}
  a = 0 , s = 0 &==> b'=0 , s'=1;\\
  a != 0 , s= 0 &==> b'=1, s' = 1 
\end{align*}
The switch and  loop (like while, for) structures can be handled similarly.

\item Variables and Data Structures: Currently the algorithm supports 
unsigned integers with small number (e.g. 4) of bits.
The fixed-size arrays and structures are flattened in the translation process. In the Integer Stack example in Figure~\ref{intstackrule} shows how an array of size $3$ is translated as 3 integer variables.
The structure elements are also flattened in the example. 
Currently our translation does not directly handle pointers and recursive data types. 
However we can manually translate the pointers into integers only if we know that the control flow of the function does not depend on the value at its pointer location.  

\item Function Calls: Currently in order to compute the abstract transition for function $f$, we inline all 
the intermediate function call inside the body of $f$.
In the guarded-update rule semantics, the rules of the intermediate functions are explicitly added
to the rules of $f$.
An explicit stack data structure is added to store the return address and the context variables.
This trick can be applied to one function calling another function as well as the  non-tail
recursive function calls.
The tail-recursive function calls can be converted into loops and do not need the stack.
In the Appendix, we show a complete translation of a recursive c function.
\end{itemize}
\section{Algorithm}

In this section we assume that the C functions are already parsed by CIL and modified into a 
software library module $Lib = (F_G, V_G, E, I)$.
We describe the basic algorithms for abstract refinement and building interface
from a given library $Lib$.
We also provide some implementation specific optimizations. 
\subsection{Basic Algorithm}
 Algorithm~\ref{explore} computes the interface for library $Lib = (F_G, V_G, E, I)$.
 The algorithm takes as input the library $Lib$, a set of functions $F \subseteq F_G$,  an abstraction $R$.
The first abstraction is obtained from the error set $E$ and initial set $I$ .
Let us define $r_1 = \set{s \in S_G \mid s \in E}$, 
$r_2 = \set{s \in S_G \mid  s \not\in E, s\in I}$ and $r_3 = \set{s \in S_G \mid s \not\in E, s \not\in I}$.
For $i \in \set{1,2,3}$, if  $r_i$ is non-empty, then we add the set to $R$ as one of the initial abstract states.
The algorithm~\ref{explore} calls AbsRef for every function $f \in F$ separately to obtain a refined abstraction $R$ w.r.t. the function.
The procedure $BuildInterface$ returns an interface graph $IG$ given the set of abstract states. 

\begin{algorithm}
\caption{Explore($Lib, F, R$)} \label{explore}
{\bf Input:}  a library $Lib = (F_G, V_G, E, I)$, set of functions $F$, abstraction $R$\\
 {\bf Output:} Interface Graph $IG$ 
 \vspace*{-1ex}
\begin{tabbing}
1.     
     \= 
  {\bf for each} $f \in F$ {\bf do}  $R { := }$ AbsRef ($R,f, E$) {\bf end for}\\
5. \>   $IG$ := BuildInterface($R, F, Lib$)\\
\end{tabbing}
\vspace*{-3ex}
\end{algorithm}
\paragraph{Modular Verification :} Each function is considered separately  in {\em AbsRef} (Algorithm~\ref{absref}).
Since, the interface graph is an input-enabled interface automata, every abstract state in the 
function can be checked separately for error reachability in one step function transition.
 The algorithm starts with the initial abstraction $R$ and the set of useful variables $V_{abs}$
 are obtained from the support set of the abstract states.
The local abstraction $R_f$ and global abstraction $R_G$ are initialized with $R$. 
The must abstraction transition is computed with respect to $R_f$ and we compute the must predecessor $S_M$ of the error set $E$.
The set $S_M$ determines the set of states of the function which eventually reach the error
set $E$.
The set $S^f_M$ is subset of $S_M$ corresponding to  the initial set of states of the function.
One-step concrete pre-image $S^1$ of $\concr{S_M}$ checks whether any new states can be 
added to $\concr{S_M}$.
If $S^1 \setm \concr{S_M}$ is non-empty then the local abstraction $R_f$ is refined and the loop
continues.
Otherwise the global abstraction $R_G$ is refined with respect to $S^f_M$.
The local and global refinements are described in the next paragraph.
The algorithm terminates when each abstract state can either reach $E$ or can not reach $E$ in one
function step. 
\begin{algorithm}[htb]
\caption{AbsRef($R,f,E$)} \label{absref}
{\bf Input:} Abstraction $R$, function $f$, error set $E$ \\
 {\bf Output:} updated $R$
 \vspace*{-1ex}
\begin{tabbing}
1.  \= $V_{abs} := \cup_{r \in R} support(r)$, $R_f {:=} R $\\
2. \>  {\bf loop}\\
3. \>  \quad $S_M := Pre^{f,R_f}_M (E)$; $S^f_M := S_M \cap I^f_L$\\
4. \>  \quad $S^1 := Pre^{f,1}(\concr{S_M})$\\
5. \>  \quad $s_{new} := S^1 \setm (\concr{S_M})$ \\
6. \>  \quad {\bf if} $s_{new} := \emptyset$ {\bf then} $R_{G} { := } R $\\ 
7. \> \quad \quad {\bf for each} $r \in R$ {\bf do}\\
8. \> \quad \quad \quad   {\bf if} $(r \cap S^f_M) \neq \emptyset\ \&\ (r \setm S^f_M) \neq \emptyset$ \\
9.  \> \quad \quad \quad \quad $R_G {:=} R_G \cup \set{r_1,r_2} \setm \set{r}$, where $r_1 := (r \cap S^f_M)$ and $r_2 :=  (r \setm S^f_M)$\\
8. \> \quad {\bf return} $R_G$\\
7. \> \quad  {\bf else} \\ 
8. \>  \quad \quad split including a variable $v$ from $\set{ v \in (V^f \setm V_{abs}) \mid v \in support(s_{new})}$\\
10. \> \quad \quad Abstraction $R_f$ is refined for all valuations of  $v$ \\
11. \> \quad {\bf end if}\\             
\end{tabbing}
\vspace*{-2ex}
\end{algorithm}
\begin{algorithm}[htb]
\caption{BuildInterface($R, F, Lib$)} \label{algo-build}
{\bf Input:} Abstraction $R$, a set of functions $F$, a library $Lib  = (F_G, V_G, E, I)$ \\
 {\bf Output:} Interface Graph $IG = (N, T, T_{e}, In, Er)$
 \vspace*{-1ex}
\begin{tabbing}
1.  \= $ Q, N,T, T_{e},In,Er = \emptyset$\\
2.  \>  $append(Q, I); append(N,I \cup E); append(In,I); append(Er,E)$\\
3.  \> {\bf while} $Q$ is non-empty {\bf do}\\
4.  \> \quad curr := removeFirst(Q)\\
5.  \> \quad {\bf for each} $f \in F$ do \\
6.  \> \quad \quad next := $Post^{f,R}_m(curr)$ \\
7.  \> \quad \quad {\bf if} ( {\bf not} member(N, next)) {\bf then} append (Q, next); append (N,next) {\bf endif}\\
8.  \> \quad \quad  {\bf if} $(next \subseteq E)$ {\bf then} $T_{e} := T_{e} \cup (curr,f,Er)$ {\bf else}  $T := T \cup (curr,f,next) {\bf endif}$\\   
9. \>  \quad {\bf end for}\\
10. \>  {\bf end while}\\  
\end{tabbing}
\vspace*{-3ex}
\end{algorithm}
\paragraph{Automatic Refinement :} 
For refinement of the local abstraction $R_f$, the algorithm finds a variable $v \in V^f$ which is not 
in the set $V_{abs}$ and is in the support set of  $S_m^1 \setm \concr{S_M}$.
The variable is added to the significant set $V_{abs}$ and a new abstraction $R^f$ is obtained
with respect to different valuations of $v$.
The refinement of global abstraction $R_G$ happens after the local abstraction reaches a fix-point
and no new states can be added in the $S_M$ set. 
For each abstract state $r \in R_G$ have a non-empty intersection with both 
$S^f_M$ and $\lnot S^f_M$, then it is split into two states $r_1$ and $r_2$.

\paragraph{Building Interface :} Algorithm~\ref{algo-build} computes the interface graph from the abstraction $R$.
For the algorithm, a list $Q$ is maintained.  
the procedure $append(Q,X)$ adds each element $x \in X$ at the end of $Q$.
The procedure $member(Q,x)$ check if $x$ is a member of $Q$.
The procedure $removeFirst (Q)$ removes the first element from $Q$ and returns the element.
The algorithm computes the next symbolic state for each element in $Q$ by applying $Post^{f,R}_m$ operator. 
There is an error- edge from the current state 
$curr$ to the error state $Er$ when the next state of $curr$ is a part of error set $E$.
Otherwise appends the next state $Q$ and a new good edge $(curr,f,next)$ is added.
The algorithm terminates when the list $Q$ is empty.

\begin{example} 
~\em
To illustrate the algorithms defined before, let us revisit the Integer Stack example (Figure~\ref{intstack}).
We assume that the guarded-update rules (Figure~\ref{intstackrule}) are converted into a 
library model with the set of functions $\set{pop, push}$.
Let us denote the state-space as $S$.
Figure~\ref{fig-run} illustrates the run of the explore algorithm(Algorithm~\ref{explore}).
The initial abstract states $r_0$, $r_1$ and $r_2$ partitions the state-space $S$ into three regions
(Figure~\ref{fig-run}(a)), where $r_0 = S\mid_{err=1}$ corresponds to error states,
$r_1 = S \mid_{err=0, top=0}$ corresponds to the initial states without error states, $r_2 =
S\mid_{err=0,top>0}$ corresponds to the non-initial non-error states. 
\begin{figure}[htb]
\centering
\includegraphics[scale=0.45]{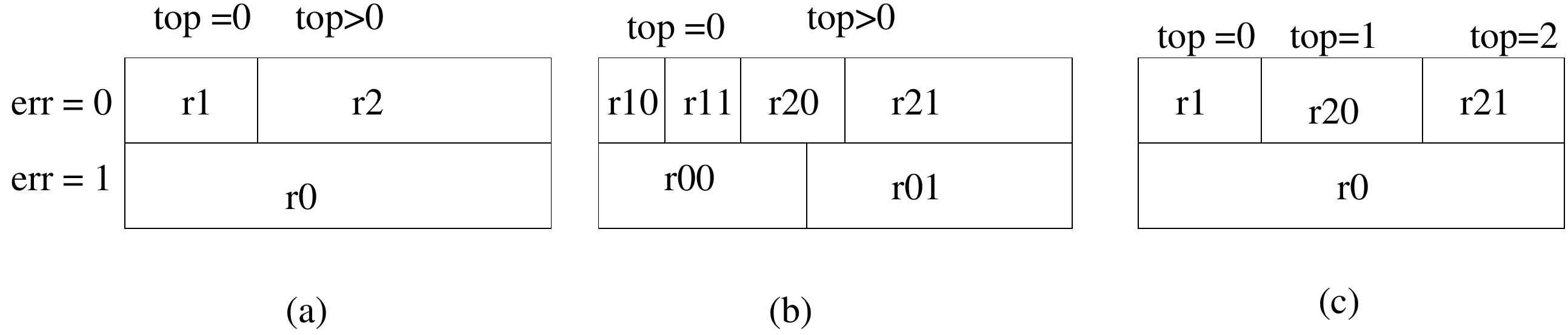}
\caption{Run of the algorithm Explore on IntStack Example. (a) The initial abstraction (b) The local
abstraction inside function (c) The final global abstraction.}
\label{fig-run}
\end{figure}
AbsRef (Algorithm~\ref{absref}) is invoked for {\em pop} function, the significant variables are
$V_{abs} := \set{err,top}$.
In the first iteration, the must predecessor $S_M$ of error state $r_0$ fail to add any new states. 
However, one step concrete predecessor of set $S_M$ returns a set  $S^1$ 
corresponding to $S\mid_{pop.s=0,top=0,err=0}$, where $pop.s$ is the local variable $s$ at function $pop$.
The support set of $S^1 \setm S_M$ contains a new variable $pop.s$ which is in $V^f$, but not  in 
$V_{abs}$. 
The local refinement of $R_f$ adds different valuations of local variable $pop.s$ (Figure~\ref{fig-run}(b)). 
The second digit of each abstract states denotes the value of $pop.s$ in the abstract state.
In the next iteration the must predecessor $S_M$ becomes $\set{r10,r00,r01}$ and no new concrete states can be added by one step predecessor of set $S_M$.
Hence the local abstraction $R_f$ can not be further refined.
The local refinement at Figure~\ref{fig-run}(b) can not be returned as as  the locally added variable $pop.s$ can not reach outside the scope of function $pop$.
The global set which leads the error set can be given by $S^f_M$ which is a subset of $S_M$ corresponding to local initial state $I^f_L$ of the pop function i.e. $S \mid_{pop.s = 0}$.
Hence the final global abstraction $R_G$ for pop function is obtained from the initial 
global abstraction $R$ of the function and will be refined with respect to set $S^f_M$ and its 
compliment set.
The algorithm returns with an unchanged global abstraction. 

Similarly for the push function  the local variable $push.s$ is included in the local abstraction.
Even if no new global variable is added in the refinement,  there is a new refinement of the global
abstract set $r_2$ with respect to the set of states (where top is 2 and err is 0) which reaches error 
states in one push call. 
The final global abstraction is shown in Figure~\ref{fig-run}(c).
The build interface algorithm (Algorithm~\ref{algo-build}) starts with the initial state $r_1$ 
and adds the edges in the graph (Figure~\ref{intstackinterface}) until every node is explored with 
respect to all functions. 
\end{example}

The interface generated by Explore algorithm is safe and permissive by construction. 
The safety in ensured by {\em AbsRef} Algorithm and permissiveness is ensured by 
{\em BuildInterface} algorithm.
The final abstraction $R$ after calling {\em AbsRef} algorithms for each function $f \in F$ distinguishes
error reaching regions from the non-reaching ones.	
In {\em BuildInterface} algorithm each function $f$ is applied in each of the states in the graph obtained
by the abstraction $R$ and hence all behaviors are captured in the interface graph.
\begin{theorem}
\label{theo-safety-permissive}
Explore (Algorithm~\ref{explore}) returns a safe and permissive interface.
\end{theorem}

\subsection{Implementation Optimizations}

\paragraph{Approximate Abstract Function Summary and Predecessors:} For practical purposes, we do not compute the abstract predecessor operators on the monolithic transition relations.
Like \cite{dAR07concur}, Equation~\ref{eq-comp-pre} holds for approximate operators.
The transition for a function $f \in F_G$ is represented as a number (say $k$) of guarded-update rules. 
For an abstraction $R \subseteq 2^{2^{S_f}}$, the must and may abstraction of rule 
$i \in \set{1,\ldots,k}$ can be given as follows:
\begin{align*}
  i.trans^{f,R}_{m+}  & := \set{ (r_1, r_2) \in (R \times R) \mid r_1 \in \absm{i.guard},\  r_2 \in \absm{i.update(\concr{r_1})}}\\
  i.trans^{f,R}_{M-}  & := \set{ (r_1, r_2) \in (R \times R) \mid r_1 \in \absM{i.guard},\ r_2 \in \absm{i.update(\concr{r_1})}}
\end{align*}
For all $j \in \set{m+,M-}$, $X \subseteq 2^R$, the approximate transition relation, one step predecessor operator and multi-step predecessor operator
can be given respectively as:
\begin{align*}
  Trans^{f,R}_{j} & := \bigcup_{i=1 \ldots k}  \   i.trans^{f,R}_{j}\\
Pre^{f,R,1}_{j}(X) & :=  \set{ r \in R \mid Trans^{f,R}_{j}(r) \cap X \neq \emptyset} \\
Pre^{f,R}_{j}\ (X) & :=  \set{ r \in R \mid r \cap (\mu Y. ( X \cup Pre^{f,R,1}_{j}(Y))) \neq \emptyset}\\
\end{align*}
.
For disjunctive transition relation, the approximate may predecessor operator will be precise; however, the approximate must predecessor will be under-approximation of the precise one.
\begin{theorem}
\label{theo-pre-approx}
For each $f \in F$, $R \subseteq 2^{2^{S_f}}$, and $X \subseteq 2^R$, we have
\[
\concr{Pre^{f,R}_{M-}(X)} \subseteq Pre^{f,*} (\concr{X}) \subseteq \concr{Pre^{f,R}_{m+}(X)}.  
\]
\end{theorem}

\paragraph{Incremental Building of Interface:} Algorithm~\ref{explore} can be used for incremental addition of function sets;  as we may not need to create the interface for all the functions at first.
The algorithm returns the refined interface for the included functions only.
The created interface can be used if we want to add more functions from the library.

\paragraph{Rule Partition for Function}
One more optimization will be partitioning the rule set of each function with respect to the abstraction
to create {\em less splitting}. 
Computation of each individual rule for must abstraction can create huge under-approximation; hence may need more splitting.
\begin{example}
In presence of If-Then-Else or Switch constructs in the source code, we may encounter the following rules after the translation. 
\begin{align*}
 r_1 : hd = true & ==> indata' =0 ; hd' = false\\
 r_2 : hd = false & ==> indata'=0 ; hd'= hd
\end{align*}
The abstract set $R$ is defined with respect to different valuations of {\em indata} variable.
If we consider each rule separately and apply the must abstraction, we miss the fact that the final value
of  variable $indata$ will be $0$ and does not depend on the initial value of $hd$.
The must predecessor of $S\mid_{indata=0}$ will be $\emptyset$ for both rules since
the must abstraction of guards will be empty-set. 
However, if we combine two rules by taking union of sets, then the must predecessor of $S\mid_{indata=0}$ will be $S$ for the combined rule and there will not be any further splitting.
\end{example}
The heuristic of rule set partition is obtained from the abstraction itself. 
If a function $f$ has $k$ rules, then $i$-th and $j$-th rules can be grouped together for an abstraction 
$R$ if  the condition  $\absm{i.guard} = \absm{j.guard}$ holds. 

\section{Results}
In this section we will provide results of some case studies and  compare with the related works.

\begin{figure}[t]
\begin{center}
\subfigure[Data Stream]{
\includegraphics[scale=0.4]{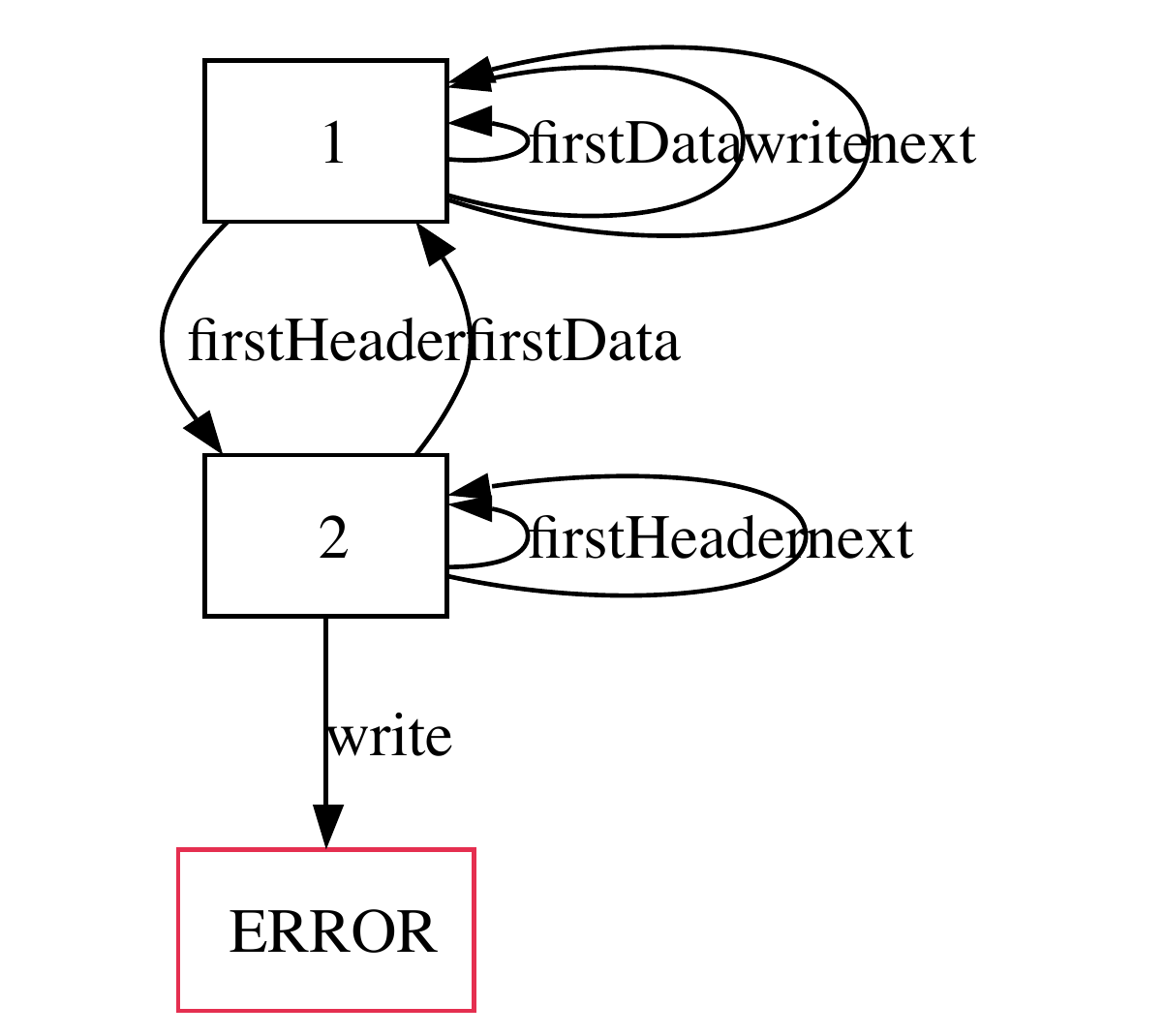}
\label{fig-interface-data}
}
\subfigure[Bit-Array-Manipulator]{
\includegraphics[scale=0.4]{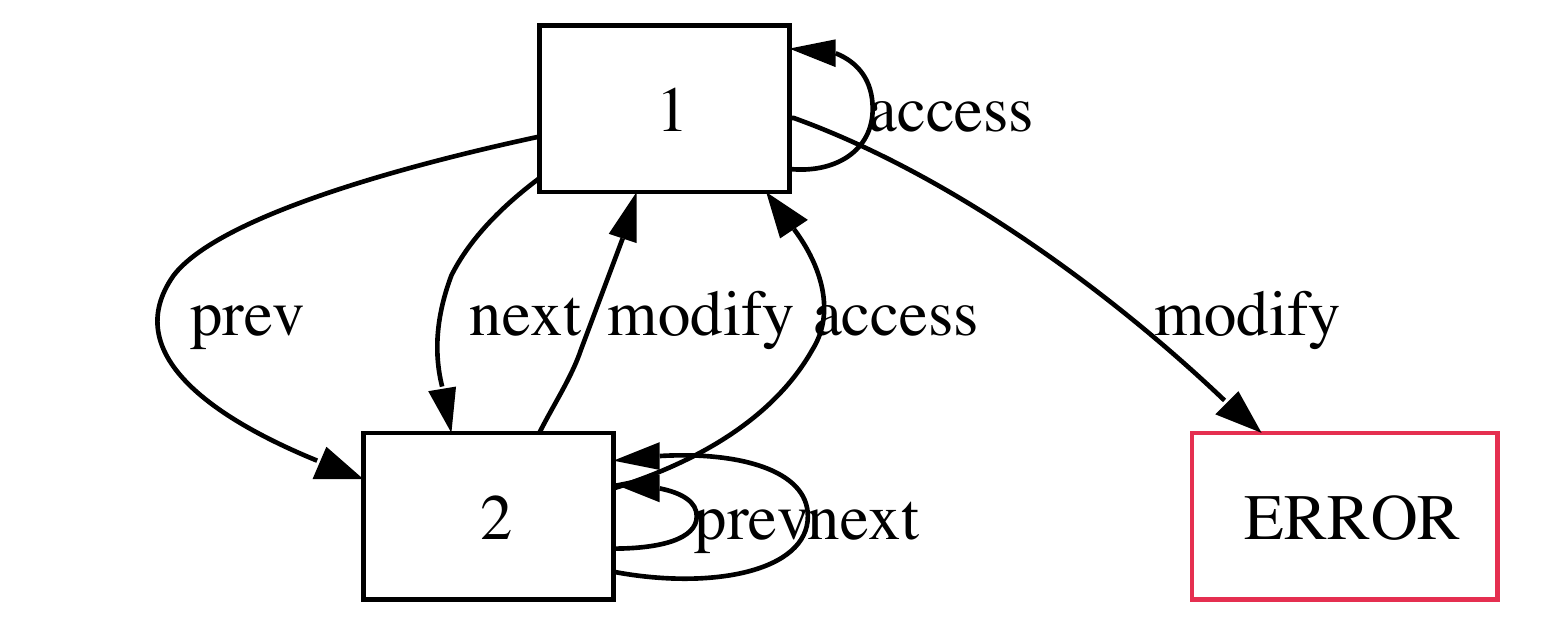}
\label{fig-interface-iterator}
}
\caption{Interfaces}
\end{center}
\end{figure}

\paragraph{Data Stream Case Study} There is a data stream with a header of length $2^h$ and data of length $2^d$ where $h \leq d$.
The program uses $d$ bits to represent the pointer and $1$ bit for the "error".
The boolean variable $isHeader$ is $1$ when in header and is $0$ otherwise. 
There are four functions in the program.
The function $FirstHeader$ and $FirstData$ takes the pointer to the first header and data location respectively. 
The function $Next$ moves the pointer within the header or data in a cyclic way.
The function $Write$ results in an error when pointer points to header section.
 Our algorithm produces the interface shown in Figure~\ref{fig-interface-data}.
 The state 1 represents that the pointer in the data part and the state 2 represents that the pointer in the header part.

\paragraph{Bit Array Manipulator} The Bit Array Manipulator has four functions : {\em prev} , {\em next},
{\em access} and {\em modify}.
Two global variables $ptr$ of length $2^k$ specify the current location of the pointer.
The global Boolean variable $valid$ denotes whether the pointer is valid.  
Another Boolean variable $err$ specify the library error states.  
The functions $next$ and $prev$ respectively increments and decrements the current pointer  and set the valid flag to true.
The functions $access$ resets the valid flag. The function $modify$ return sets $err$ to true when the valid is false, otherwise sets valid to false. 
Our algorithm produces the interface shown in Figure~\ref{fig-interface-iterator}.
The state 1 represents that the valid bit is false and the state 2 represents that the valid bit is true.


\begin{figure}[htb]
\begin{tabular}{| l | l| l l |l l l|}
\hline
Case Study\ & \ Params\ \ &\  Time (ms)\ &\ Regions\ &\ Direct\ &\ Learning\ &\ CEGAR\ \\
\hline
Data Stream & $h=2,d=12$ &  3  & 2 & 1028 & 2 & 257\\
                              & $h=4,d=12$ &  4  & 2 & 4112 & 2 & 257\\
                              & $h=13, d=13$ &  18 & 2 & 16384 & 2 & 2\\ 
\hline 
Bit Array           &  $k=8$  & 2 &  2 & 68 & 2 & 2\\
Manipulator     &   $k=9$  & 4  &   2 & 130 & 2 & 2\\
                           &   $k=16$  & 8  &   2 & 16386 & Timeout & 2\\
\hline 
\end{tabular}
\label{results}
\caption{Results}\label{tab-res}
\end{figure}

\paragraph{Comparison}  Figure~\ref{tab-res} shows a comparison of our algorithm with the related work on these two examples.
The first two columns show the name and different parameter values of the case-studies.
The next column describes the running time (in milli seconds) of explore algorithm from the parsed guarded-update rules.  
The next column represent the number of non-error regions in the interface graph.
The non-error regions from other three related work are given in the last three columns and the data is obtained from Beyer et. al.'s. work~\cite{beyerTR06}.
The results for Direct algorithm show that direct algorithm runs fastest, but the size of interface graph is exponential in $d$.
We obtain that the CEGAR algorithm provides minimal graph only when $h=d$ in the Data Stream example.
The size of the graph in the CEGAR algorithm depends on the proper representation of variables with Boolean variables.
The CEGAR approach refine by adding a new boolean variable; which has a risk of splitting many abstract states unnecessarily. 
In contrast, our algorithm keeps global abstraction separate from local abstraction inside the function and 
refines the global abstraction lazily with respect to the final reachable set ($S^f_M$). 
Learning algorithm provides the minimal graph, but slowest of all three approaches.
Our algorithm provides the same number of non-error regions as the learning algorithm. 
However, we can not compare time due to different platforms.

\section{Application of Interfaces}

In this section, we show how a safe and permissive interface can be useful in the verification and testing of the software programs.
The following section briefly describe the modifications needed for the interface to be compatible with these settings. 

\subsection{Software Verification with Interfaces}

Let us assume that we have computed an interface graph for a set of functions.
Given a client program consisting of those functions one can immediately check the client with respect to the interface graph.
The idea would be simulating the actions of the client program into the interface graph and check whether the library error state (State "ERROR") is reached.
For example, a client with a single line $modify(b)$ on the BitArrayManipulator $b$ can be simulated in the interface graph 
(Figure~\ref{fig-interface-iterator}).
We can see that the error state ERROR is reached from the initial state (State 1).
There could be an infinite number of possible clients corresponding to those functions and each of them
can be model-checked after the interface is computed.  

\subsection{Offline Test Case Generation}
In the model-based testing paradigm, an implementation under test (IUT) is checked with respect to a given model program 
(a specification of the IUT).
Our algorithm can build an interface graph from the definitions of the functions given in the model program.
We can create a C source regression test-suite from the interface generated from the libraries. 
However, we need to extend the function calls with the argument values to create a test-bench for the IUT.
For example, Figure\ref{intstackcode} can be generated from the model program in Figure\ref{intstackinterface}. 
If we are given a linked-list  implementation of a finite-size integer stack,  we can create an offline test-suite 
from the interface graph.
The testing of the implementation with respect to the test-suite checks whether the interface goes to the error state if and only
if the implementation goes to the error state.
If there is a discrepancy between the behavior of the interface graph and the code, we understand the implementation source
needs further checking.

\section{Conclusions}
In this section we conclude with the summary of the work and possible future directions.
We have provided a new algorithm for interface synthesis with a  local-global abstraction refinement 
framework.
This framework is can dramatically reduce the state-space of the interface generation by hiding
local variables inside each function.
The abstract summarization of the functions provides scalability. 
The modular analysis is used to handle each function separately.
In our generalized setting any C-style set of functions can be handled.

The results show that our algorithm provides a safe, permissive and sufficiently minimal (i.e. comparable to the learning algorithms) interface from the set of functions.
 We have provided the approximate abstract predecessor operators to handle the state-space inside
 the function.
 The interface synthesis can be incremental : hence one can add new functions to the interface 
 and it may lead to refinements corresponding to the function.

 The interface could be used to immediately verify clients and  as offline test-suite for a new untested  implementation.
 However, the translation engine is very basic and some parts are done manually.
In future we like to work more on covering more aspects (e.g. pointers, recursive data types) of the 
C source code such that we can have bigger case studies.
We like to see how we can use the shape analysis algorithms to translate complex data types.
We also like to include CIL inside the tool TICC s.t. it can parse C functions and represent the rules directly in MDD format.
We like to implement the back-end using a combination of MDD and SMT solvers such that the 
space-space problems can be handled better.

 \bibliographystyle{abbrv}
 \bibliography{dvlab}
\section*{Appendix}
A C function to compute n-th Fibonacci number is translated into  a set of guard-update rules.
To handle the activation stack and store the context of the caller, there is an explicit implementation
of integer stack. The variable {\em nextpc} denotes the next value of the location variable
after return from one of the the stack operations.
The variable $v$ contains value of input parameter of push and  is assigned before a call to push
. $v$ is the output parameter of pop and obtained after returns from pop.
{\small
\begin{verbatim}
module Fibonacci:
   var i,s,top : [0..MAX]
   var v:[0..15]
   var a0, a1, .... : [0..15]
   var nextpc: [0..31] 
output push: {   
       s=15 & top < MAX ==> top'=top+1 &  i'=top & s'=16; 
       s=16 & i=0 ==> s'=nextpc & a0'=v;
       .............  
       }
output pop :{
      s=17 ==> i'=top & t'=18;
      s=18 & i=0 ==> s'=19 & v' = a0;
      ...........
      s=19 & i>0 ==> top'=i-1 & s' = nextpc
}
...
endmodule
\end{verbatim}
}
\noindent
The rule set {\em fib} defines the transitions inside the Fibonacci function. 
The variable $res$ stores the result when the call returns and $tmp1$ and $tmp2$ are two 
temporary variables.  
A recursive call to itself is translated into saving the return address,  the current value of n,
initializing n for the called function and a subsequent jump to the initial location of the function. 
{\footnotesize
\begin{verbatim}
var n : [0...20]      
var res, tmp1, tmp2 : [0..31]                       
output fib: {                                                 
  s=0 & n<3 ==>  res'=1 & s'=11;             
  s=0 & n>=3 ==> s'=2;                               
  s=2 ==> nextpc' = 3 & s'=15 & v'=5;          
  s=3 ==> nextpc' = 4 & s'=15 & v' =n;       
  s=4 ==> n' = n -1 & s'=0;                           
  s=5 ==> t'=6 & tmp1' = res;                      
  s=6 ==> nextpc' = 7 & s'=15 & v'=9;          
  s=7 ==> nextpc' = 8 & s'=15 & v'=n;        
  s=8 ==> n'=n-2 & s'=0;                                
  s=9 ==> s'=10 & tmp2'= res;                    
  s=10 ==> s'=11 & res' = tmp1+tmp2;         
  s=11 ==> nextpc' = 12 & s'=17;                    
  s=12 ==> n' = v & s'=13;                           
  s=13 ==> nextpc' = 14 & s'=15;               
  s=14 ==> s' = v;                                           
}                                                                         
\end{verbatim}
}

\end{document}